\PassOptionsToClass{natbib=true}{acmart}
\documentclass[sigconf]{acmart}
\AtBeginDocument{%
  }

\copyrightyear{2026}
\acmYear{2026}
\acmConference[SCA/HPCAsia 2026]{SCA/HPCAsia 2026: Supercomputing Asia and International Conference on High Performance Computing in Asia Pacific Region}{January 26--29, 2026}{Osaka, Japan}
\acmBooktitle{SCA/HPCAsia 2026: Supercomputing Asia and International Conference on High Performance Computing in Asia Pacific Region (SCA/HPCAsia 2026), January 26--29, 2026, Osaka, Japan}
\acmDOI{10.1145/3773656.3773669}
\acmISBN{979-8-4007-2067-3/26/01}

\makeatletter %
\renewcommand{\@authornotemark}{}%
\makeatother

\usepackage{csquotes}
\usepackage{fontawesome5}

\begin{document}

\title[The Munich Quantum Software Stack]{The Munich Quantum Software Stack}
\subtitle{Connecting End Users, Integrating Diverse Quantum Technologies, Accelerating HPC}

\author{
  Lukas Burgholzer\textsuperscript{$\dagger$ $\ddagger$ *},
  Jorge Echavarria\textsuperscript{$\S$ *},
  Patrick Hopf\textsuperscript{$\dagger$ $\ddagger$},
  Yannick Stade\textsuperscript{$\dagger$},
  Damian Rovara\textsuperscript{$\dagger$},\\
  Ludwig Schmid\textsuperscript{$\dagger$},
  Ercüment Kaya\textsuperscript{$\dagger$ $\S$},
  Burak Mete\textsuperscript{$\dagger$ $\S$},
  Muhammad Nufail Farooqi\textsuperscript{$\S$},
  Minh Chung\textsuperscript{$\S$},\\
  Marco De Pascale\textsuperscript{$\S$},
  Laura Schulz\textsuperscript{$\P$},
  Martin Schulz\textsuperscript{$\dagger$ $\S$},
  Robert Wille\textsuperscript{$\dagger$ $\ddagger$ $\|$}
}
\authornote{Both authors contributed equally to this work.}
\affiliation{
  \institution{\vspace{0.5em} \textsuperscript{$\dagger$}Technical University of Munich, Munich, Bavaria, Germany}
  \institution{\textsuperscript{$\ddagger$}Munich Quantum Software Company (MQSC), Garching near Munich, Bavaria, Germany}
  \institution{\textsuperscript{$\|$}Software Competence Center Hagenberg, Hagenberg, Upper Austria, Austria}
  \institution{\textsuperscript{$\S$}Leibniz Supercomputing Centre, Garching, Bavaria, Germany}
  \institution{\textsuperscript{$\P$}Argonne National Laboratory, Lemont, Illinois, USA}
  \vspace{0.5em}
  \institution{\{lukas.burgholzer, patrick.hopf, yannick.stade, damian.rovara, ludwig.schmid, \\ ercument.kaya, burak.mete, martin.schulz, robert.wille\}@tum.de}
  \institution{\{jorge.echavarria, muhammad.farooqi, minh.chung, marco.pascale\}@lrz.de}
  \institution{lschulz@anl.gov\vspace{13pt}}
  \city{}
  \country{}
}
\authorsaddresses{}

\renewcommand{\shortauthors}{Burgholzer, Echavarria, et al.}

\begin{abstract}
    Quantum computing is advancing rapidly, with notable progress in both hardware and algorithm development.
    However, to make quantum computing truly accessible and usable across disciplines, a comprehensive, efficient and unified software stack is essential.
    Such a software stack must be flexible enough to support diverse hardware platforms and evolving algorithms, while enabling programming models for a wide range of users from quantum specialists to users without deep quantum expertise.
    Additionally, a software stack must support dynamic resource management for efficient system utilization as well as seamless integration with classical High-Performance Computing (HPC) environments. The latter is key, as quantum systems are increasingly envisioned as accelerators within hybrid workflows, supporting various levels of integration from loosely to tightly coupled.
    Despite numerous discussions, conceptual proposals, and initial efforts in recent years, few tangible implementations of such full-featured software stacks currently exist.

    This paper introduces the Munich Quantum Software Stack (MQSS)---a modular, open-source, and community-driven ecosystem designed to support hybrid quantum-classical workflows and applications.
    We present the multi-layered architecture of MQSS, which enables a seamless execution of high-level user applications on heterogeneous quantum hardware back-ends, while also supporting the coupling with the corresponding classical workloads.
    Key components include front-end adapters for popular frameworks as well as new programming approaches, an HPC-integrated scheduler, a powerful MLIR-based compiler infrastructure, and a standardized hardware abstraction layer known as the \mbox{Quantum Device Management Interface (QDMI)}.
    
    \noindent While still under active development, the MQSS already now offers sophisticated concepts and open-source implementations that lay the foundation for a robust quantum computing software stack, with a forward-looking design that also considers the future requirements of fault-tolerant quantum computing, such as support for various qubit encodings and mid-circuit measurements.
    More information on the MQSS is available at
    \begin{itemize} %
    \item[\faGithub] \small{\url{https://github.com/Munich-Quantum-Software-Stack}}
    \item[\faBook] \small{\href{https://www.munich-quantum-valley.de/research/research-areas/mqss} %
                               {https://www.munich-quantum-valley.de/research/ \newline research-areas/mqss}}
    \end{itemize}

\end{abstract}


\maketitle

\begin{figure*}[!h]
    \begin{minipage}[t]{0.49\textwidth}
        \begin{minipage}[b]{\textwidth}
            \centering
            \includegraphics[width=\textwidth]{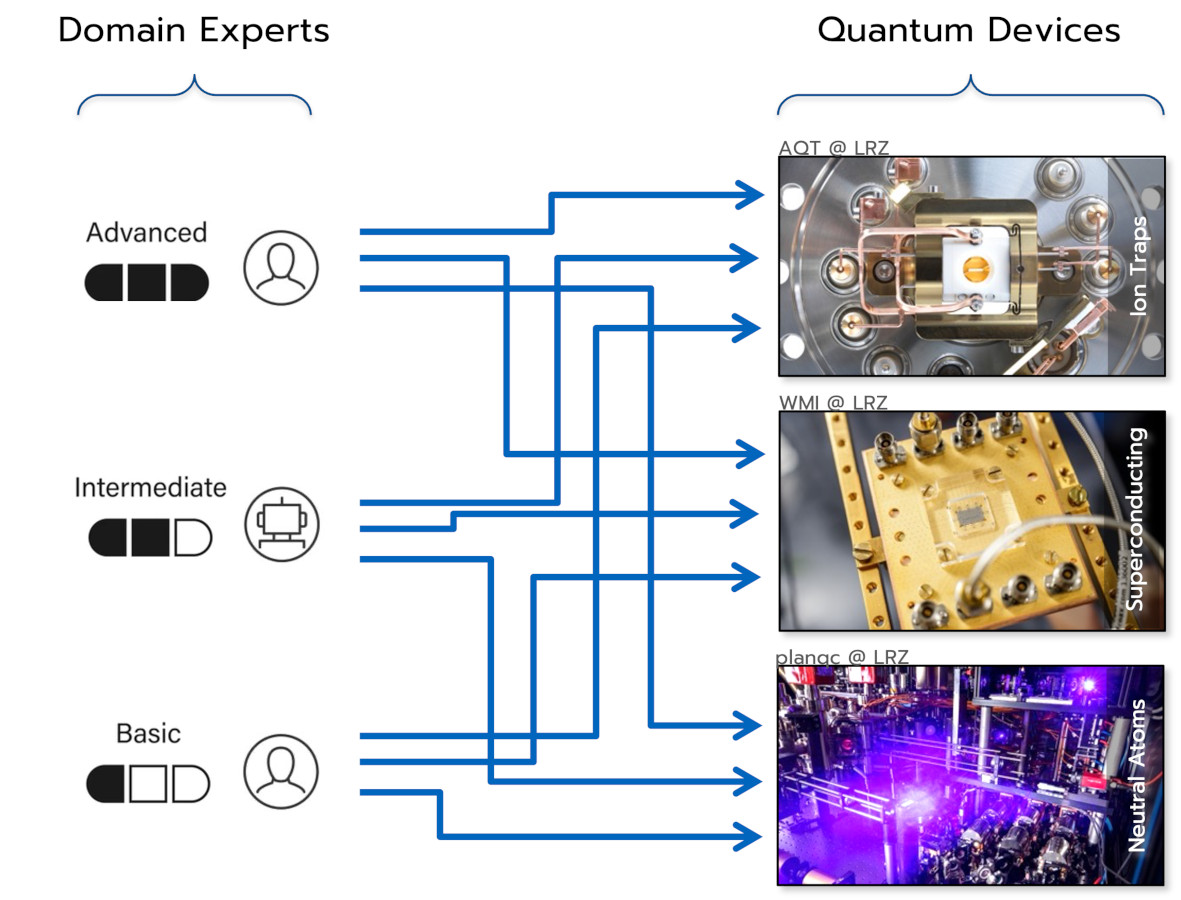}\par
            (a)
        \end{minipage}
    \end{minipage}\hfill
    \begin{minipage}[t]{0.49\textwidth}
        \begin{minipage}[b]{\textwidth}
            \centering
            \includegraphics[width=\textwidth]{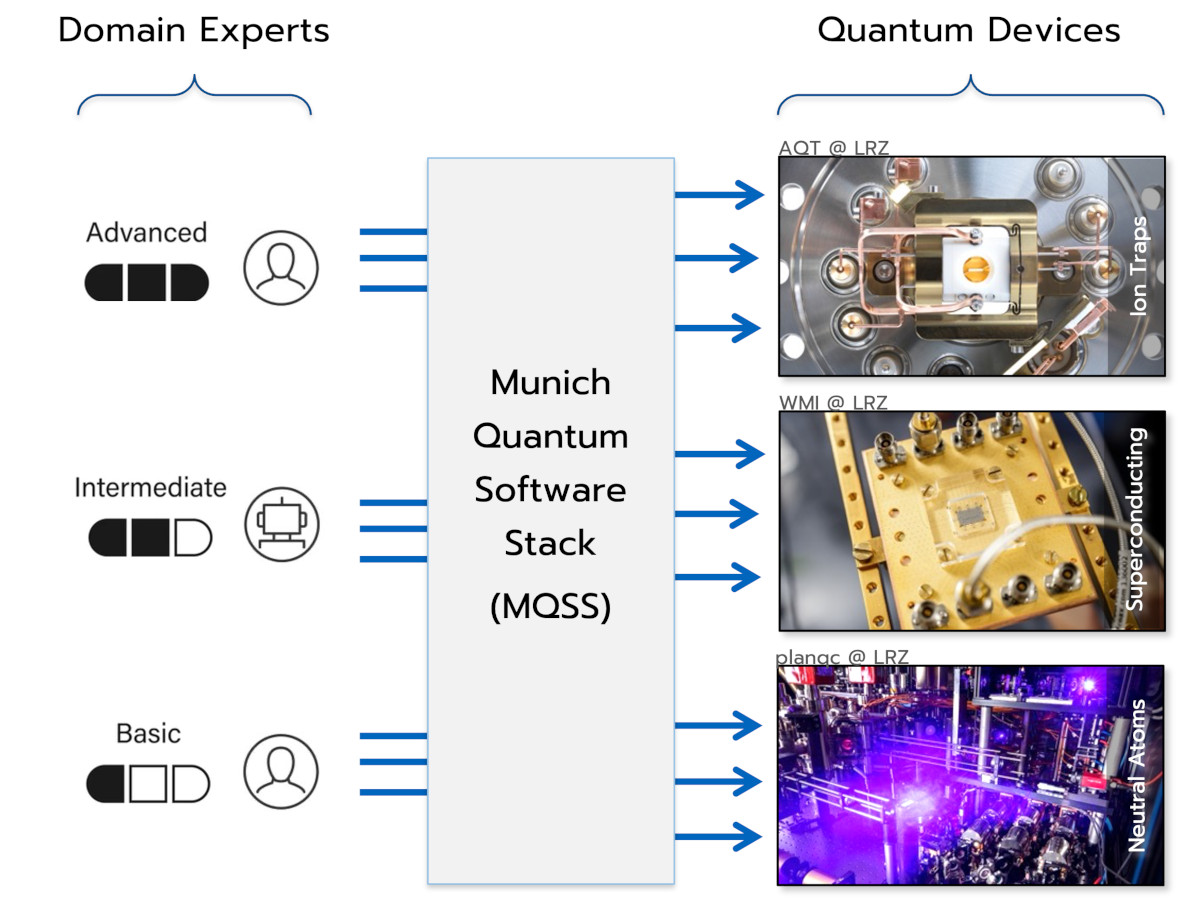}
            (b)
        \end{minipage}
    \end{minipage}%
    \caption{a) Traditional view: many users, many systems, many stacks.  b) Vision going forward: decoupling front-end and back-end in a single stack via shared IRs.}
    \Description{Two side-by-side diagrams comparing quantum software stack approaches. Left diagram (a) shows multiple separate stacks connecting different user types to different quantum technologies. Right diagram (b) shows a unified MQSS stack connecting all user types to all quantum technologies.}
    \label{sfig:testb}
\end{figure*}

\section{Introduction}\label{sec:introduction}

Quantum Computing (QC) promises to revolutionize computation by tackling problems intractable for even the most powerful classical supercomputers.
Particularly in the realm of High-Performance Computing (HPC), quantum computers are envisioned as specialized accelerators, similar to Graphics Processing Units (GPUs) or Field Programmable Gate Arrays (FPGAs), that can efficiently solve specific computationally hard sub-problems in science and industry.
However, realizing this potential requires more than just building quantum hardware.
It requires a concerted and collaborative effort across a wide spectrum of disciplines.
No single research group or company can overcome the multifaceted challenges alone.
Instead, a successful quantum ecosystem relies on the synergy of: 1) experimental physicists and electrical engineers who provide the hardware; 2) mathematicians and theoreticians who lay the theoretical foundations for quantum algorithms and error correction; 3) computer scientists who design and implement the needed hardware and software environments; 4) software engineers who maintain and support the required software ecosystem suitable for production usage; and 5) domain experts from industry and science who identify and formulate compelling use cases.

Only in such a synergetic environment, can we create the needed sophisticated software stack, as the critical link that connects the diverse domains and stakeholders.
Its role is to translate high-level problem descriptions from domain experts into low-level, hardware-specific instructions for the quantum computer.
Furthermore, it must seamlessly integrate Quantum Processing Units (QPUs) into existing classical HPC infrastructures (a concept referred to as HPCQC integration), thereby enabling the hybrid quantum-classical workflows that are expected to dominate the field for the foreseeable future.

The necessity of such a software stack is underscored by the emergence of large-scale quantum initiatives worldwide.
While none of these initiatives are primarily focused on software, they all recognize the importance of a robust software ecosystem.
The Munich Quantum Valley (MQV), for example, is such a large-scale initiative that brings together over 400 researchers and practitioners from academia, industry, and government to create a comprehensive quantum ecosystem, where the software stack is a core pillar alongside hardware development.

This paper introduces the Munich Quantum Software Stack (MQSS), an open-source, community-driven effort to create such a software ecosystem.
The MQSS is designed to be a modular, efficient, and extensible platform that bridges the gap between high-level quantum applications and the intricacies of diverse quantum hardware back-ends.
It provides a comprehensive set of tools for compilation, optimization, and execution of quantum circuits, while also facilitating the deep integration with classical HPC systems for hybrid workflows and applications, which are required for practical quantum advantage.
The MQSS's capabilities were first demonstrated when it successfully achieved HPCQC tightly-coupled integration, a first between on-premise HPC and QC systems.
To achieve this, SuperMUC-NG---the flagship HPC system at the Leibniz Supercomputing Centre (LRZ), one of Germany's three national HPC centers---was coupled with Q-Exa, Germany's first quantum demonstrator: a 20-qubit IQM Radiance superconducting device. This integration enabled the execution of a VQE application, resulting in one of the world's first HPC-quantum computing (HPCQC) demonstrations~\cite{bickley2025extendingquantumcomputingsubspace}.

\subsection{The Need and Challenges for a Quantum Software Stack}\label{sec:why-stack}

In classical computing, we typically expect and rely on a mature and reliable software stack as an essential component of the ecosystem, providing layers of abstraction that allow developers to productively use complex hardware\footnote{Typically in the form of an operating system environment, such as Linux or Windows. Another example is the E4S stack stemming from the US Exascale Computing Program (ECP), see \url{https://e4s.io}.}.
The same necessity holds for quantum computing, and even more so in the context of hybrid HPCQC, where quantum processors are envisioned to function as accelerators, analogous to GPUs or FPGAs.
A combined software stack, with shared components across both ecosystems, is the critical infrastructure that enables this vision of hybrid quantum-classical computing~\cite{shehataBuildingSoftwareStack2025, elsharkawyIntegrationQuantumAccelerators2025, seitzUnifiedHybridHPCQC2023, saurabhConceptualArchitectureQuantumHPC2023}.
Moreover, users, both humans and automated tools, differ in expertise, requirements, and capabilities.
As shown in \autoref{sfig:testb}.a, currently, many existing stacks address these user types separately for each quantum technology.
\autoref{sfig:testb}.b, on the other hand, shows the MQSS's unified solution that supports all user domain–technology permutations.

A stack of this nature must be modular, flexible as well as efficient.
It needs to support a diverse and evolving landscape of quantum hardware, from superconducting circuits to trapped ions, from neutral atoms to photonic systems, and more~\cite{muraliArchitectingNoisyIntermediateScale2020, haeffnerQuantumComputingTrapped2008,bluvsteinArchitecturalMechanismsUniversal2025,obrienPhotonicQuantumTechnologies2009}.
Each of these modalities has its own unique characteristics, constraints, and performance metrics, which must be abstracted away to provide consistent access.
The stack must also be capable of accommodating a wide range of current and future quantum algorithms, as well as hybrid applications that combine the usage of classical and quantum resources.
To make quantum resources accessible to users who are not quantum physicists, the stack must offer high-level abstraction layers, covering the complex workflow for given problem descriptions to their execution on a back-end.
Further, quantum computing adds new requirements, such as dynamic Just-In-Time (JIT) compilation, as the software stack must adapt to real-time changes in the hardware's state as well as performance characteristics to optimize execution.

The architectural principles of such a stack must be flexible enough to support a wide range of deployment models from on-premise setups and cloud access to QCs.
Only this will allow for a broad adoption and a reuse of compiler components, programming models and back-end solutions.
In this paper, though, we focus on a deep integration with on-premise HPC systems.
For one, this represents one of the extreme cases stressing the software stack the most, and, second, it enables tightly-coupled hybrid computations, which are expected to have the most immediate impact for real-world use cases.
This integration is key to unlocking the full potential of quantum accelerators, moving beyond simple, loosely-coupled cloud access.

Finally, the software stack must be designed with the future in mind, providing a clear path from today's Noisy Intermediate-Scale Quantum (NISQ~\cite{preskillQuantumComputingNISQ2018}) devices to tomorrow's fault-tolerant machines~\cite{shorFaulttolerantQuantumComputation1996,ryan-andersonImplementingFaulttolerantEntangling2022,eganFaulttolerantControlErrorcorrected2021}.
This will require flexibility on representing encodings, supporting mid-circuit measurements, enabling back-ends with hardware support, as well as supporting bi-directional state transformation up and down the stack.

The development of a comprehensive quantum software stack is fraught with challenges that stem from the unique nature of quantum computing and its ecosystem.
As a relatively new field, quantum computing lacks the established standards and deep-rooted abstractions that we naturally rely on in classical computing.
The underlying hardware is incredibly diverse, with various physical realizations of qubits---such as superconducting circuits, trapped ions, neutral atoms, or photons---each presenting distinct properties, connectivity, and error characteristics~\cite{devoretImplementingQubitsSuperconducting2004,haeffnerQuantumComputingTrapped2008,schmidComputationalCapabilitiesCompiler2024,stadeAbstractModelEfficient2024}.
Even systems based on the same underlying qubit technology can vary wildly in their architecture and performance.
This heterogeneity makes it difficult to create a unified software layer that can effectively target all platforms.

Furthermore, the quantum computing community is a melting pot of different expertise.
Software development has often been driven by physicists focused on the underlying quantum systems, rather than by computer scientists with expertise in building scalable and maintainable software.
On the user side, domain experts from fields like chemistry or finance want to solve their specific problems, but are often neither experts in quantum computing nor in HPC.
Conversely, HPC experts may lack a deep understanding of quantum mechanics, and quantum computing experts may not be familiar with the intricacies of large-scale classical resource management.
These conditions create significant communication gaps between stakeholders, which need to be bridged by a concerted effort to establish a common language and shared understanding.
This is essential for creating tools that are both powerful and usable.

\subsection{Our Approach: The Munich Quantum Software Stack}\label{sec:contribution}
The MQSS is our answer to these challenges.
Developed by computer scientists, but in close connection with all other stakeholders and their expertise, it provides both a conceptual blueprint, which covers the fundamental design of hybrid quantum/classic software stacks, and a tangible, open-source software stack with a concrete implementation that can be deployed to and used by a wide range of users, systems, and deployment environments.

Our approach is centered on several key principles that directly address the challenges outlined above:
\begin{itemize}
    \item To tackle hardware \textbf{heterogeneity}, the MQSS features a \textbf{modular architecture} designed for extensibility and flexibility, allowing for the seamless integration of new components and support for a growing number of quantum platforms.
    \item To bridge the gap with classical computing, it is built from the ground up for \textbf{deep integration with classical HPC}, enabling the tightly-coupled hybrid workflows that are essential for quantum acceleration.
    \item To foster a collaborative ecosystem and overcome the lack of standards, we develop and promote \textbf{open interfaces and specifications}, such as the Quantum Device Management Interface (QDMI~\cite{willeQDMIQuantumDevice2024}).
\end{itemize}
This commitment to openness and interoperability stands in contrast to other, more monolithic, single-vendor stacks and ensures that the MQSS can serve as a versatile and future-proof platform for the entire quantum community, empowering users and developers alike.
These principles are not just abstract concepts, but concrete solutions that guide the development of the MQSS, ensuring it is a robust, usable, and impactful tool for advancing quantum computing.

The MQSS is actively being developed and deployed, supported with Continuous Integration (CI) and Continuous Deployment (CD) techniques, and its components are successively being released as open source and integrated with external components.
This makes the MQSS a community-driven effort that has the ability to bring together a wide range of users, vendors, centers, and compiler developers, which all can base their work on the open platform and then contribute components back to the overall ecosystem.

\section{Architectural Overview}\label{sec:overview}

This section provides a high-level overview of the MQSS. We begin by outlining the intended use cases and target user groups of the stack to illustrate the problems it aims to solve.
We then describe the overall architecture of the MQSS, detailing its main components and how they interact to form a cohesive and powerful software ecosystem for HPCQC integration.
This overview sets the stage for a more detailed look at the individual components in the following section.

\subsection{Use Cases and Target Groups}\label{sec:use-cases}
The MQSS is designed to serve a diverse set of users and enable a wide range of applications and use cases, bridging the gap between the abstract potential of quantum computing and its practical implementation within an HPC context.
Our primary goal is to make quantum resources accessible and useful to several key target groups:
\begin{itemize}
    \item \textbf{Domain Experts} from fields such as material science, chemistry, operational research, or finance who possess deep knowledge of their problem domain, but are not necessarily experts in quantum computing. For them, the MQSS provides high-level abstractions that allow them to formulate their problems without needing to manage the complexity of the underlying quantum hardware.
    \item \textbf{HPC Experts and Users} who are accustomed to heterogeneous computing environments and wish to leverage quantum computers as another type of specialized accelerator, much like a GPU or an FPGA. The MQSS facilitates this by integrating QPUs as offload devices with classical HPC resources, enabling various levels of hybrid workflows from loosely to tightly-coupled~\cite{elsharkawyIntegrationQuantumAccelerators2025}.
    \item \textbf{Quantum Algorithm and Software Developers} who require a flexible, modular platform to design, test, and benchmark new algorithms, compilation strategies, and software tools. The open-source and extensible nature of the MQSS provides an ideal environment for this kind of research and development.
    \item \textbf{Hardware Providers and Physicists} who develop and operate quantum computing hardware. For this group, the MQSS offers (quasi-)standardized interfaces, like QDMI, for integrating new devices into the ecosystem, allowing them to expose their systems to a broader user base and receive valuable feedback. At the same, it also enables low-level access through the same infrastructure to facilitate low-level experiments and manual access.
    \item \textbf{System Administrators and User Managers} who are responsible for operating QC or HPCQC systems. For this group, the MQSS offers introspection interfaces to understand system health and status, as well as interfaces to integrate issues, like authentication and quota management, with the matching HPC components.
\end{itemize}

The MQSS supports this wide range of users and use cases, merging the access into a single unified infrastructure.
Further, it does so by not \enquote{just} supporting the execution of quantum programs, but by also offering a native integration with existing classical systems, in particular High Performance Computing (HPC) systems.
This is critical for the success of any quantum ecosystem, because QC systems alone cannot support themselves, as they need classical systems for staging and control, and because they cannot support arbitrary workloads, as quantum algorithms only target very specific computational problems.

As a consequence, QC system will for the foreseeable future operate as quantum accelerators, where a QPU tackles a computationally hard subtask or kernel of a larger problem that, in its whole, still continues to be orchestrated by a classical HPC system.
The result is a powerful hybrid workflow exploiting and combining the respective strengths from both worlds.

\begin{figure*}[htbp]
  \vspace{15pt}
  \includegraphics[width=\textwidth]{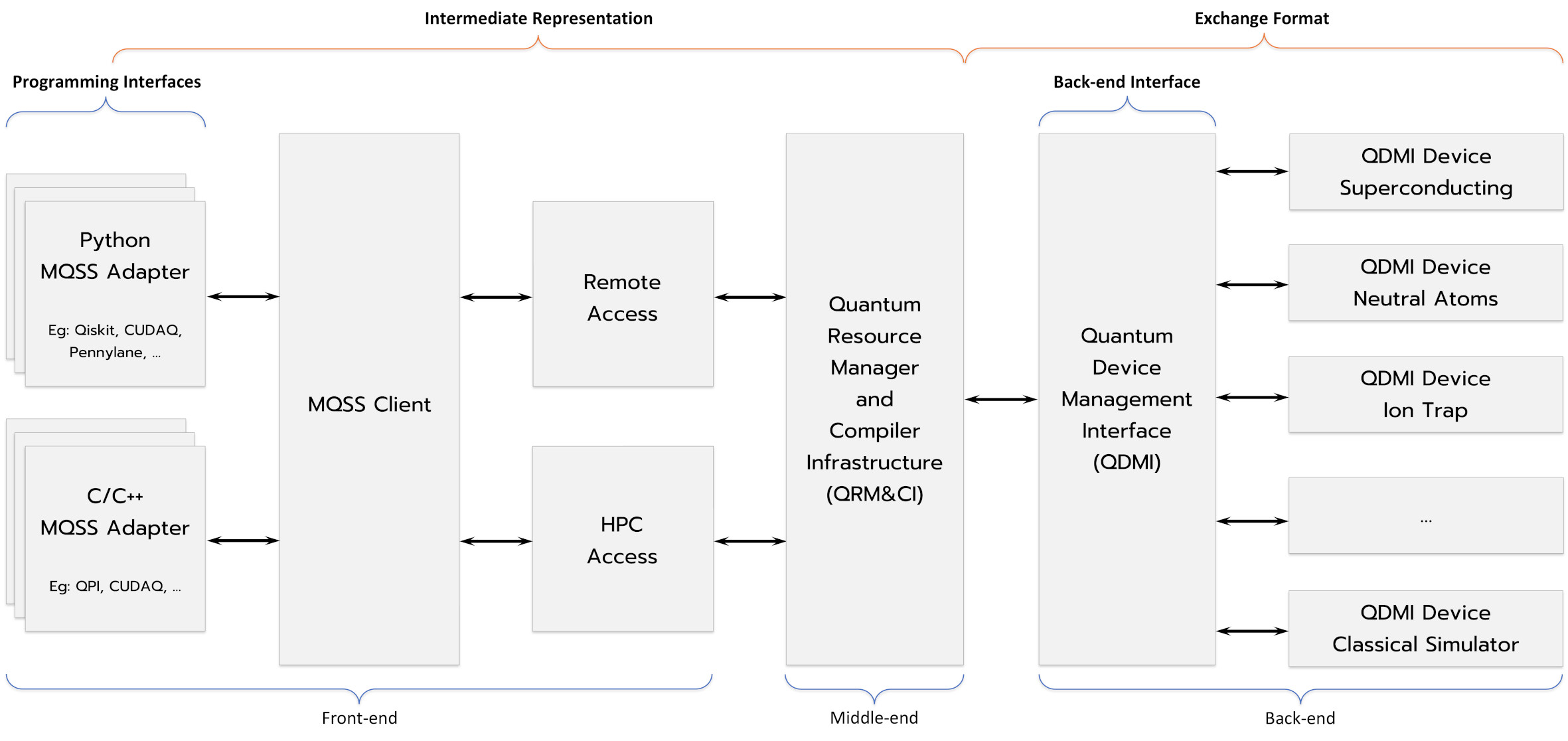}
  \caption{Overview of the MQSS and its components, connecting end-users and their high-level problem descriptions (left) to a diverse set of quantum hardware back-ends, as well as simulators (right).}
  \Description{Layout showing the MQSS architecture from left to right: users and applications on the left connect through front-end adapters and the MQSS client to the middle-end layer containing the compiler and resource manager, which then connects via QDMI to various quantum back-ends including superconducting, trapped-ion, and neutral atom systems on the right.}
  \label{fig:overview-figure}
\vspace{10pt}
\end{figure*}

Executing such a hybrid workflow, though, involves a complex interplay of data flow, resource management, and scheduling.
For example, a user's quantum program must be compiled for a specific device, the device must be available, and the classical and quantum computations must be synchronized., while at the same time minimizing idle times on both the QC and HPC system.

A concrete instance of the MQSS is deployed at the Leibniz Supercomputing Centre (LRZ), which currently hosts many of the quantum systems available via the MQV.
This deployment, which marks the first HPCQC integration in Germany, as well as one of the first co-located on-premise integration world-wide, demonstrates our stack's versatility in a production environment.

The MQSS offers multiple access paths---including a web portal (the \enquote{MQSS Dashboard}), command-line access, and tightly integrated hybrid access from within the HPC environments.

\subsection{The MQSS Core Architecture}\label{sec:architecture-overview}
To realize the use cases described above, the MQSS is built as a modular, multi-layered architecture, implemented as a series of connected components. \autoref{fig:overview-figure} sketches the basic architecture, forming the bridge from users (left) to systems (right).

At its heart, the MQSS consists of a common and agnostic infrastructure, built on top of established open source principles, which can then be instantiated in many scenarios and customized via plugins operating on standardized interfaces.
This design philosophy emphasizes flexibility and extensibility, allowing components to be replaced or upgraded and enabling many-to-many connections between the layers.

The key architectural layers, which are explicitly designed to support the different user groups, include:
\begin{itemize}
    \item \textbf{Front-end Layer:} This is the highest level of abstraction, providing user-facing interfaces and decoupling them from the underlying runtime system. It includes support for popular quantum programming frameworks, like Qiskit and PennyLane, domain-specific tools that allow experts to formulate problems in a familiar language, and new programmatic interfaces, like the Quantum Programming Interface (QPI;~\cite{kayaQPIProgrammingInterface2024}) for hybrid environments.
    \item \textbf{Middle-end Layer:} This layer is responsible for embedding the quantum workflows into the classical HPC environment. It handles the scheduling of hybrid jobs, manages the allocation of quantum resources, and orchestrates the communication between them. It also contains the powerful and flexible compiler infrastructure that translates high-level quantum circuit descriptions into executable instructions optimized for a specific hardware back-end.
    \item \textbf{Back-end Layer:} This layer provides a unified interface to the underlying quantum hardware. Through a dedicated interface, called QDMI, it abstracts access to the specific details of each back-end, allowing the higher levels of the stack to interact with different QPUs in a standardized way. It also exposes crucial information about the hardware, such as calibration data and Figures of Merit and Constraints (referred to as FoMaCs in the remainder of the text), which are essential for hardware-aware compilation and optimization~\cite{Hopf_2025}.
\end{itemize}
This layered architecture ensures that the MQSS can evolve with the rapidly changing quantum landscape, providing a stable, yet adaptable platform for HPCQC integration.

\begin{figure*}[t!]
    \centering
    \includegraphics[width=0.375\textwidth]{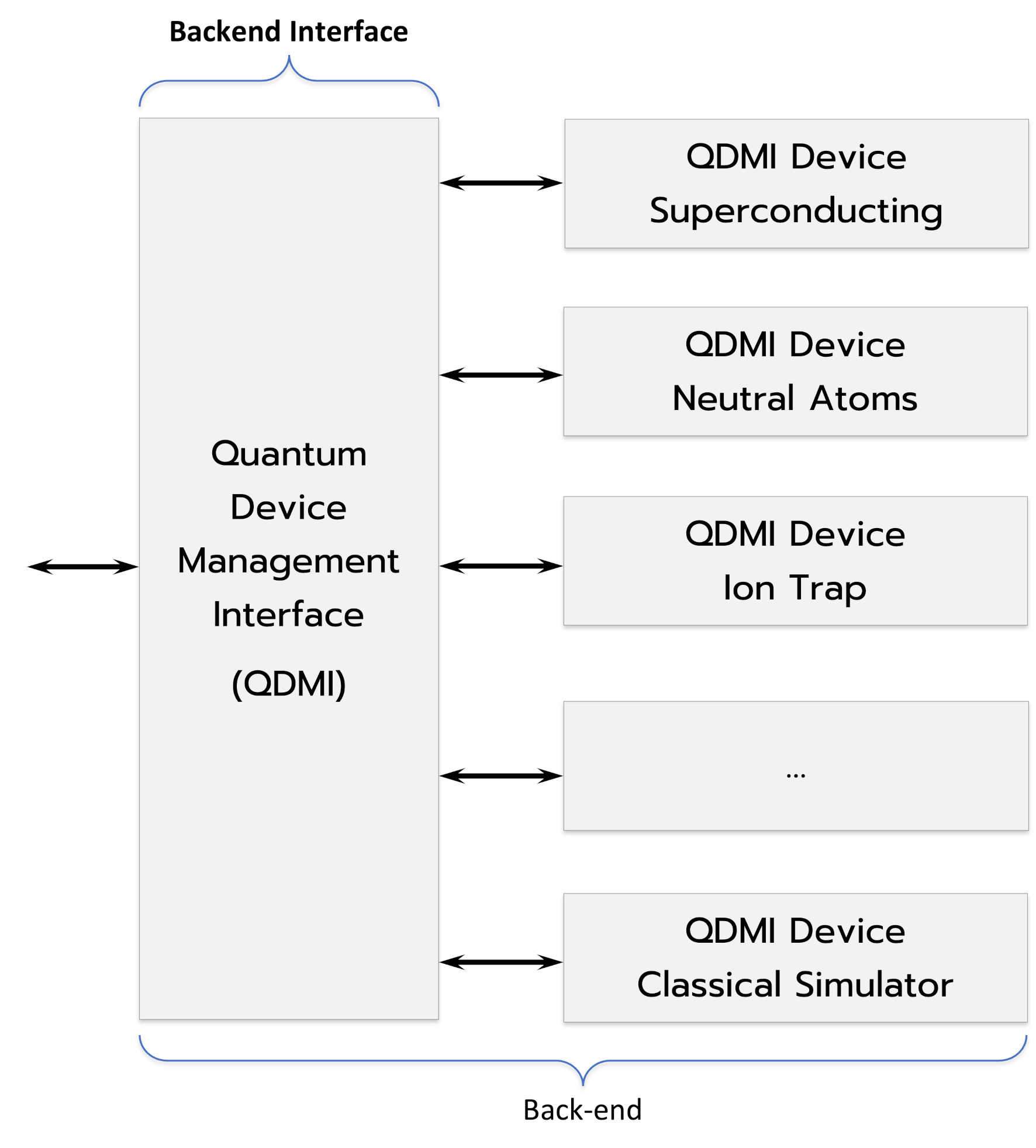}
    \caption{Structure of the MQSS back-end, featuring the implementation of the QDMI interface as well as a series of system device plugins.}
    \Description{Layered diagram showing the MQSS back-end structure with QDMI at the top providing a unified interface, connected to multiple device plugins below, each interfacing with different quantum hardware systems from various vendors.}
    \label{fig:MQSS-back-end}
\end{figure*}

\newpage
\section{Individual Components of the Software Stack}\label{sec:components}

This section details the key components of the MQSS, looking at the stack bottom up, adding layers of abstraction on top of the hardware systems up to the point where the MQSS directly supports domain users and their applications.

At the bottom of the stack, the MQSS features a unified interface for running individual quantum kernels (using gate or pulse notation~\cite{Echavarria_Pulse}), delivering results, and querying both static and dynamic information of QC systems and their physical environment.
The latter information is then aggregated into higher-level data in the form of Figures of Merit and Constraints (FoMaCs), which summarize the status, the health, and the system configuration.
This data is then fed to other layers of the stack to influence and optimize compilation and scheduling decisions, as well as to inform users and administrators.

Building on this, we introduce the Quantum Resource Manager \& Compiler Infrastructure (QRM\&CI) that handles the entire compilation, resource management and scheduling.

Finally, we present the front-end adapters, which provide various levels of abstraction, from familiar---typically low-level---SDKs, like Qiskit, to hybrid HPCQC languages, offload approaches and high-level Domain Specific Languages (DSLs) and libraries.
These enable users to leverage the full power of the underlying stack, based on their application requirements and prior knowledge.

\newpage
The following sections summarize the main concepts in these layers.
Links to more details and actual (open-source) implementations are provided where available for a deeper understanding.

\subsection{Quantum Device Management Interface}\label{sec:qdmi}

As depicted in \autoref{fig:MQSS-back-end}, the lowest layer of the software stack interfaces directly with the hardware devices and their control systems.
The level of abstraction needed at that level can depend on the targeted device, its control structure and internal logic, its type of access (e.g., gate or pulse-based), and its ability to execute code itself.
This can reach from directly accessing the control logic to managing autonomous devices, e.g., in the form of control processors with conditional logic, with their own complex firmware stack.

\begin{figure*}[htbp]
    \vspace{10pt}
    \centering
     \includegraphics[width=0.8\textwidth]{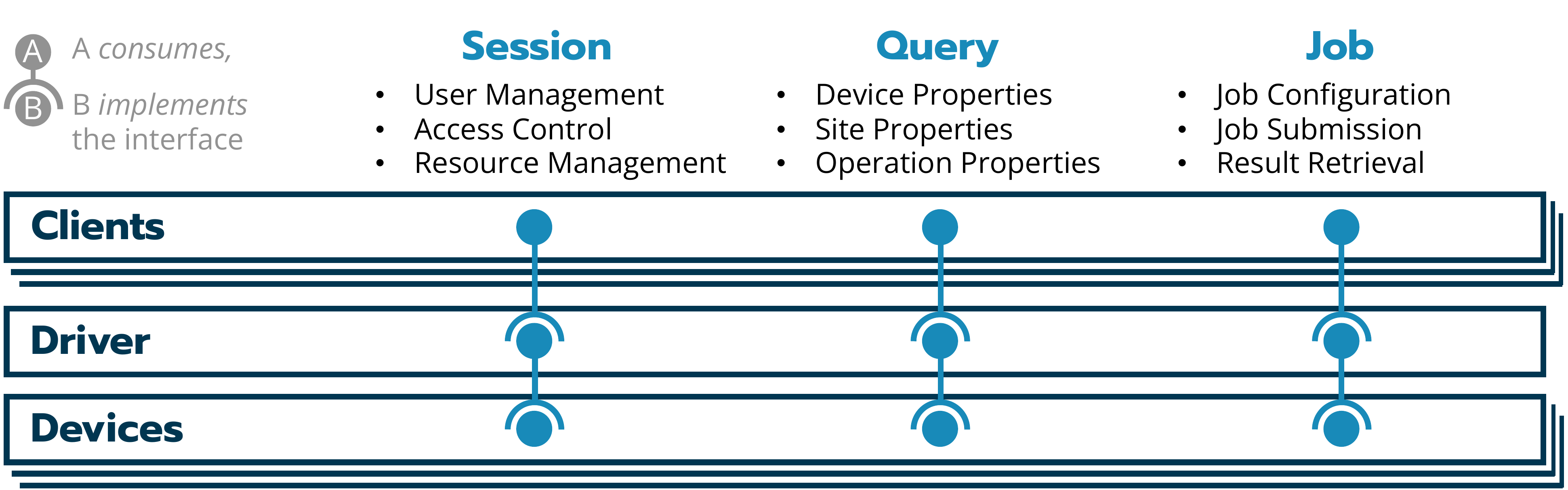}
    \caption{The three main components of the QDMI: Session Management, Job Submission, and Query Interface.}
    \Description{Schematic diagram illustrating QDMI's three main functional components: Session Management for handling multi-user and multi-device access, Job Submission for sending and managing quantum kernels, and Query Interface for retrieving device properties and calibration data.}
    \label{fig:qdmi-prongs}
    \vspace{10pt}
\end{figure*}

The Quantum Device Management Interface (QDMI) is designed to bridge these differences and offers a standardized connection to QC devices covering multiple access options and supporting different types of firmware stacks.
It is split into three main components (see \autoref{fig:qdmi-prongs}): 1) a session management enables the software stack to work with multiple users on multiple devices in a safe and reliable manner, 2) a job submission and management interface to send jobs (in one of many forms) to a device, and 3) a query interface to extract device properties and constraints, e.g., live fidelity values or concrete gate sets supported by the device.

Its primary role is to abstract the hardware-specific details of device control and data acquisition, enabling the submission and management of quantum kernels in a uniform manner.
Kernels are represented in an exchange format, which is produced directly by the compiler, as explained below, and then handed on to the respective back-end for final transformation on the vendor side (if needed) and execution.
For the latter, QDMI defines a simple queue management (incl. submission, asynchronous progress, and job status queries) associated with clearly defined and separated sessions, enabling the execution of multiple kernels from multiple users on multiple devices.

Additionally, QDMI allows software tools to retrieve device properties.
This includes static information, like the number of qubits or available gate sets, as well as dynamic information, like gate fidelities, calibration data, or data from environmental sensors.
This information can then be analyzed and used by compilers, the underlying runtime or the user to understand system properties and adjust their usage accordingly.

On the implementation side, QDMI can be split into two main parts: the interface specification describing the interface towards both the user and the hardware vendor, and a core implementation that implements the session, job and query management.
This implementation can then be customized with plugins, each representing the needed control software for individual or families of quantum devices.
This structure allows the software stack to access the devices via a firmly defined stack-facing interface, while vendors can add support for QDMI by implementing a thin plugin based on the hardware-facing interface.
This approach not only supports easy expandability, but also supports portability across software stacks and quantum technologies.

Looking forward, this modular and portable approach opens QDMI for further extensions.
In particular, it easily supports different levels of abstraction from gate-based to pulse-based circuits, it enables both directly controlling execution to code shipping to \enquote{close to metal} inside the quantum device, all with the same interface.
Further, it will be possible to integrate virtual devices, external software stacks, as well as ISV software, or remote execution on devices, without changes to the interface.

For more information on QDMI, including its specification and implementation, please refer to:
\begin{itemize} %
    \item[\faGithub] \small{\url{https://github.com/Munich-Quantum-Software-Stack/QDMI}}
    \item[\faBook] \small{\url{https://munich-quantum-software-stack.github.io/QDMI}}
\end{itemize}

\subsection{Quantum Resource Manager \& Compiler Infrastructure}\label{sec:qrmc}
\begin{figure*}[t!]
    \vspace{10pt}
    \centering
    \includegraphics[width=0.375\textwidth]{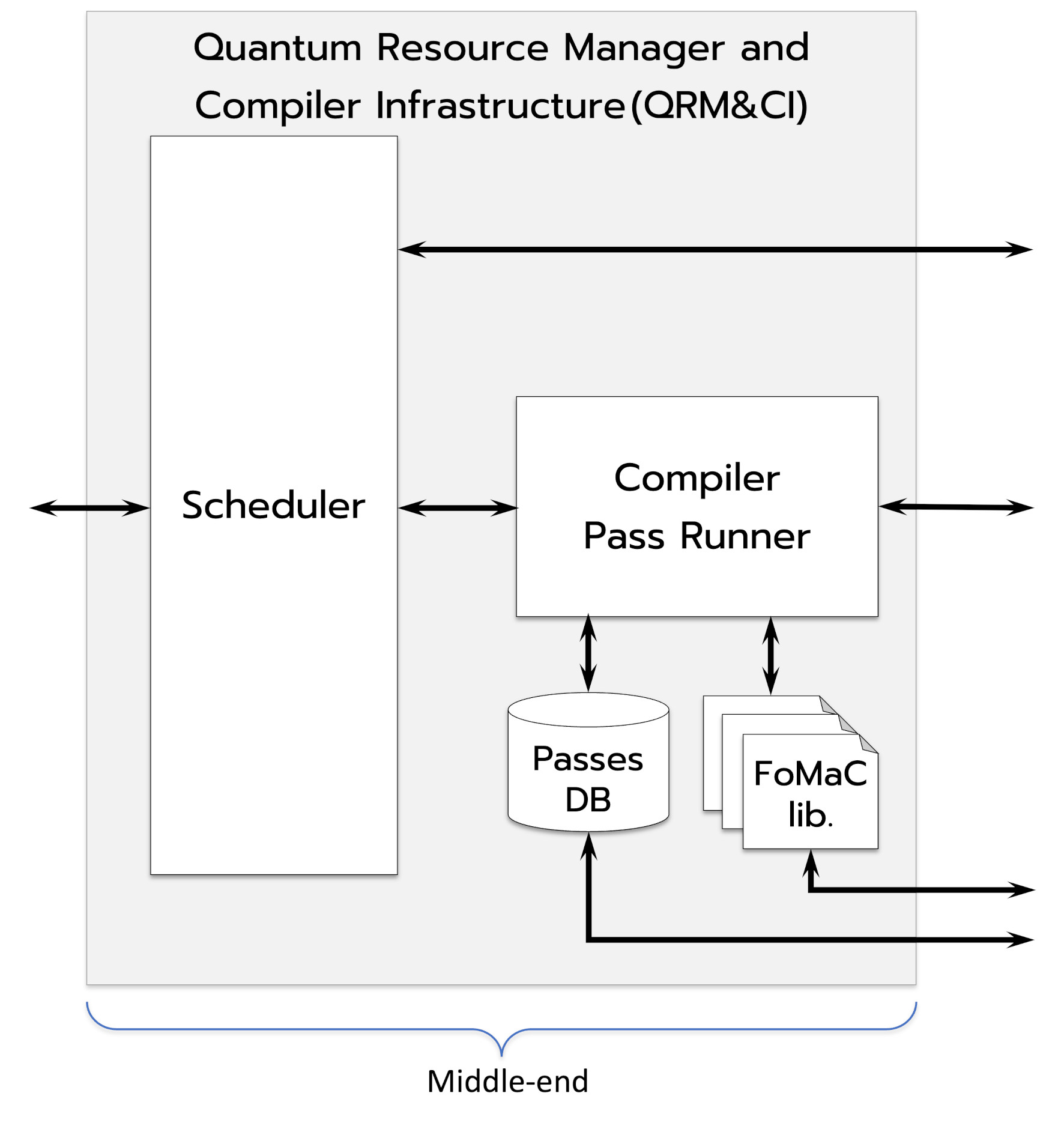}
    \caption{The structure of the MQSS middle-end, featuring the QRM\&CI and its internal components implementing the compiler.}
    \Description{Block diagram of the MQSS middle-end showing the Quantum Resource Manager and Compiler Infrastructure with its key components: FoMaC support libraries for hardware metrics, the MLIR-based compiler with device-agnostic and device-specific passes, and the scheduler for resource management.}
    \label{fig:MQSS-middle-end}
    \vspace{10pt}
\end{figure*}

The Quantum Resource Manager \& Compiler Infrastructure (QRM\&CI), shown in \autoref{fig:MQSS-middle-end}, is the central runtime component responsible for orchestrating the execution of quantum kernels.
It combines a dynamic compiler infrastructure with scheduling and resource management, acting as the bridge between user input in the form of quantum kernels on one side and execution on the quantum systems on the other.
It receives kernels in some, likely high-level, Intermediate Representation (IR), transforms and lowers them to execution-ready kernels in a, likely low-level and system-specific IR, allocates the needed quantum resources, and then uses QDMI to execute the kernel.
Once complete, it reads the result, again via QDMI, transforms the results into the form requested by the user, and returns it via the initial programming abstraction.

The QRM\&CI is composed of several key components, including support libraries for Figures of Merit and Constraints (FoMaCs), a compiler infrastructure, and a scheduler, which are detailed in the following subsections.

\subsubsection{Figures of Merit and Constraints Support Libraries}\label{sec:fomac}

While QDMI provides the  interface to query raw hardware and system properties, this information is low-level and hard to use directly.
It must first be analyzed and refined, a task that every consumer of the data needs and hence should not be replicated.
This task is, therefore, relegated to a set of libraries, which use the QDMI-provided raw data and create concrete Figures of Merit and Constraints (FoMaCs).
This is particularly helpful for complex, dynamic, and time-sensitive information, such as gate fidelities, coherence times, and readout error rates, which are aggregated from the low-level data provided by the hardware.

Such live, aggregated telemetry is critical for the hardware-aware compilation and optimization and, hence, is made available by a set of FoMaC libraries to the rest of the MQSS software stack.
For example, the compiler can use up-to-date fidelity data to choose the best physical qubits for a computation or to adjust its optimization strategy to avoid particularly noisy gates.
The FoMaC libraries also play a role in monitoring the broader operational environment.
They can aggregate data from environmental sensors---provided through QDMI---that track factors like temperature and magnetic fields, which can influence the performance of sensitive quantum hardware.
By providing this comprehensive and dynamic view of the hardware's status, FoMaC libraries enable the software stack to make intelligent, data-driven decisions that maximize the performance and reliability of the targeted quantum systems.

A simple example implementation of a FoMaC library is provided as part of the QDMI repository.
A more complex example is the sys-sage library, which can be used to capture, analyze and track topology and reliability data from quantum systems~\cite{mishraTowardsUnifiedArchitectural2025}.
sys-sage was originally developed for HPC systems to track classic node topologies and performance properties, and was extended to include quantum data, offering a unified view of HPC and quantum system information.
This demonstrates how the MQSS supports the integration of the two technologies.

\subsubsection{Compiler Infrastructure}\label{sec:mlir}
One of the key elements of the QRM\&CI is its compilation infrastructure (in the following referred to as compiler), which translates input kernels to low-level, hardware-specific sequences of instructions, which can be consumed by QDMI and its back-end plugins.
For this, the compiler uses a family of Intermediate Representations (IR) to describe the quantum circuits and applies a series of compilation passes to these kernels to transform them step by step and gradually lower them from a likely high-level and generic input circuit description to likely low-level and execution-ready code.
In this process, each pass can apply optimizations for the circuits, be guided by the FoMaC input, verify properties of circuits for debugging, select appropriate devices based on circuit properties, or transpile circuits to the available native gate sets, as queried via QDMI.

To manage this complexity, to support a compatible family of IRs, to maintain flexibility, and to reuse existing work, the MQSS leverages the Multi-Level Intermediate Representation (MLIR) framework~\cite{mlir}, developed for and widely used in classical systems.
MLIR allows for the creation of a modular and extensible compiler pipeline where different levels of abstraction are represented by distinct IR \enquote{dialects}.
The framework further supports extensibility via custom dialects and transformation passes through its plugin system~\cite{sca_hpcasia_2026_2}.
This allows the MQSS to be both back-end-agnostic at its higher levels and highly specific and optimized at its lower levels, all the way to low-level representations, like QIR~\cite{QIRSpec2021, Stade_QIR}, while at the same time supporting low-level intrinsics and annotations where helpful in high-level abstractions.
It also allows the combination of dialects of IRs for classical and quantum kernels, enabling a truly hybrid compilation and cross-paradigm optimization.
Further, it supports different levels of abstractions on the back-end side, easing the integration of new back-ends.

The compiler is designed as part of the runtime system to be highly
dynamic, using the live data provided by the FoMaC support libraries (see \autoref{sec:fomac}) to inform its optimization choices.
The actual passes are picked from a library of passes based on the matching IRs, the targeted systems, as well as user directives, similar to
compiler optimization techniques from classical computing.
To further support pass selection and optimization, the MQSS can also employ AI-based pass selection, which can be integrated via a separate pass selection interface.
Tools, like the MQT Predictor tool~\cite{quetschlichMQTPredictorAutomatic2024}, can use this to select a dedicated compilation pass sequence for a given circuit based on its properties and the current state of the available hardware.

The modular structure of the MQSS compiler and the distinction between a generic infrastructure with a customizable pass selection mechanism, coupled with an extensive and expandable pass library (with MQSS and external contributions) offers the ability to grow and extend the MQSS capabilities, as well as to reuse existing techniques and software.
Specifically, the MQSS leverages existing passes from frameworks such as Qiskit, but also incorporates individually developed passes for superconducting systems~\cite{willeMappingQuantumCircuits2019,zulehnerEfficientMethodologyMapping2019,zou2024lightsabrelightweightenhancedsabre,liuTacklingQubitMapping2023}, neutral atoms~\cite{stadeAbstractModelEfficient2024,stadeOptimalStatePreparation2025}, and trapped ions~\cite{schoenbergerShuttlingScalableTrappedIon2024,vandeweteringConstructingQuantumCircuits2021}, such as those available from the Munich Quantum Toolkit (MQT)~\cite{willeMQTHandbookSummary2024, Burgholzer2025}.

For more information on the MQSS compiler infrastructure, including a collection of passes, please refer to:
\begin{itemize} %
    \item[\faGithub] \small{\href{https://github.com/Munich-Quantum-Software-Stack/MQSS-Passes-Suite} %
                                 {https://github.com/Munich-Quantum-Software-Stack/ \newline MQSS-Passes-Suite}}
\end{itemize}

\begin{figure*}[t!]
    \vspace{10pt}
    \centering
    \includegraphics[width=0.57\textwidth]{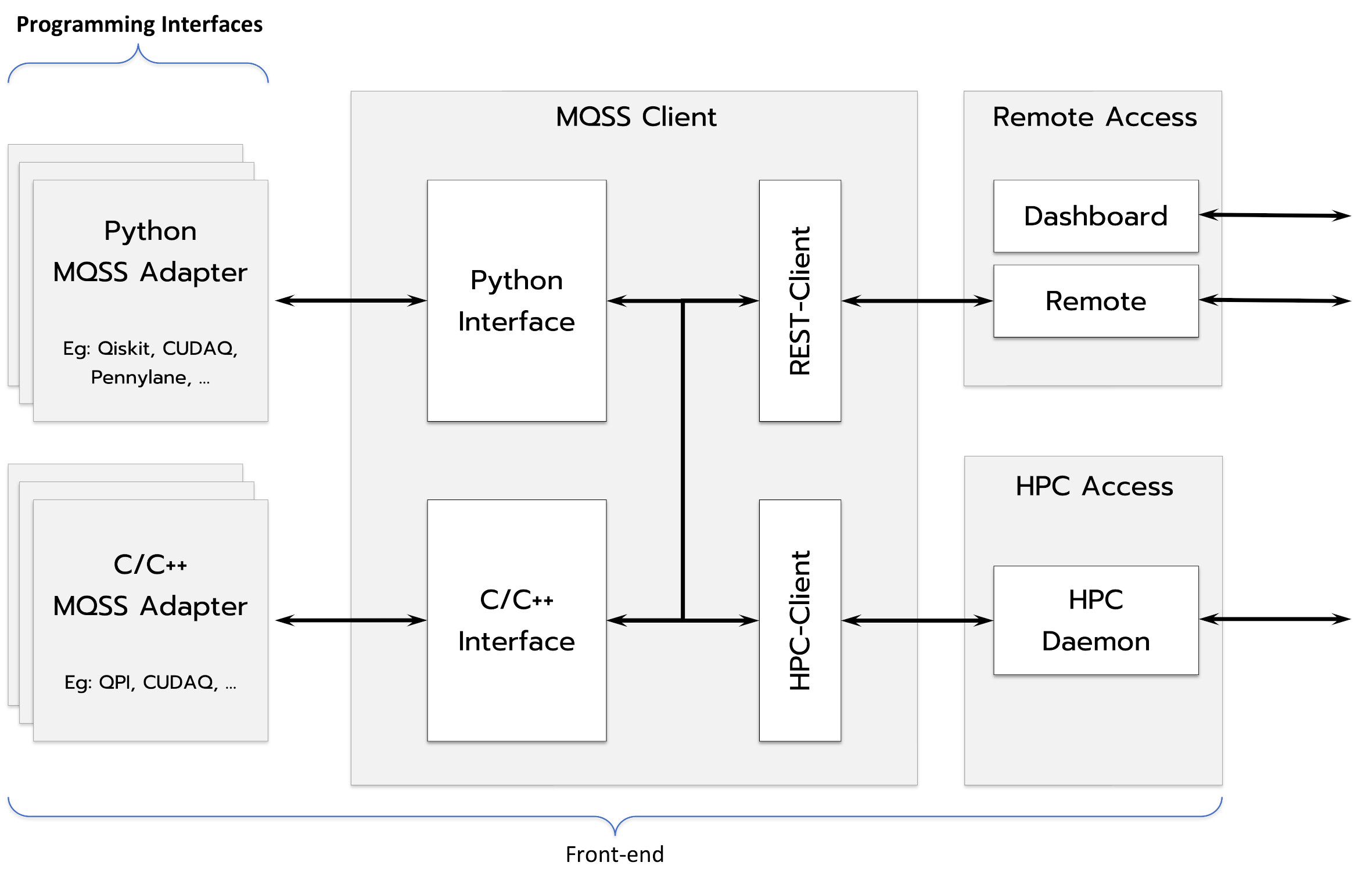}
    \caption{Structure of the MQSS front-end, featuring the MQSS Client connecting end-user programming to the rest of the stack.}
    \Description{Diagram showing the MQSS front-end layer with multiple adapters for different programming frameworks including Qiskit, PennyLane, CUDA-Q, and QPI on the left, all connecting through the central MQSS Client to the rest of the software stack on the right.}
    \label{fig:MQSS-front-end}
    \vspace{10pt}
\end{figure*}

\subsubsection{Scheduling and Resource Management}\label{sec:scheduling}
To effectively execute quantum circuits on quantum systems, which are in most cases a rare resource, a software stack must include the needed resource management.
This will enable the user and/or the system to select an appropriate QC system and then schedule a kernel to that system in a way that both reduces latency and improves throughput.
Both aspects are potentially highly dependent on kernel and system properties, as well as the current live state of the system.

Such a scheduler will naturally have to be bi-level.
At the upper level, the QRM\&CI selects a device for execution and then coordinates execution with simple execution and shot management inside the QDMI back-ends.
While the latter is kept explicitly simple and typically only uses single priority queue for each device, the former scheduler is intended to be more sophisticated.
Through explicit interfaces, plugins can be loaded to support adaptive device selection, e.g., using AI methods~\cite{quetschlichMQTPredictorAutomatic2024} as well as performance prediction to estimate workload and system occupancy.

\subsubsection{HPC Integration}\label{sec:hpcqc}

For the integration into classical systems, in particular into HPC environments with their own resource management and scheduler, the topic of resource management becomes even more critical.
Here, the QRM\&CI scheduler additionally has to coordinate with the HPC scheduler, such as Slurm~\cite{Slurm}, that manages the HPC component of the hybrid workload, resulting a three-level scheduling system.
The goal of this scheduler is to minimize idle times for both the HPC and the QC system, while at the same time selecting the most suited quantum system for the supplied kernel.

In order to fulfill these requirements, the QRM\&CI requires a sophisticated scheduling policy as well as tight coordination with the HPC scheduler, the quantum scheduler, as well as the system status via QDMI. Further, it must understand the characteristics of the input circuit as they would map to a specific technology with its respective native gate set, fidelity and execution speed.
Based on that, it must be able to predict execution times to understand resource requirements and length of reservations, so kernel executions can be scheduled, possibly across jobs and users, in a way that the QC resource is available when needed, but used otherwise when not.
This requires the ability to model and predict the performance of both the runtime stack and the actual quantum execution in the respective back-end.
For this, the MQSS offers the needed interfaces to add and to customize such models, and with that to optimize overall system utilization.

\subsection{Front-end Adapters and MQSS Client}\label{sec:frontend}

Facing the user and/or the application, the MQSS Adapters, as shown on the left side of \autoref{fig:MQSS-front-end}, provide the primary entry point for defining and executing quantum kernels, either via direct coding or via automated tools.
These adapters, therefore, allow users to leverage the entire stack using a wide range of APIs, abstraction libraries, DSLs or programming models. From a technical point of view, they decouple the programming model from the underlying compilation and runtime stack, a concept well-known from classical computation, but so far not widely used for quantum programming.

Recognizing that users as well as existing tools have diverse backgrounds and prefer different programming paradigms, the MQSS is designed to be compatible with a wide range of popular quantum software development kits and interfaces, each connected with its own, lightweight front-end adapter, but connecting to the same stack (see \autoref{sfig:testb}).
This compatibility ensures that developers can continue to work in the environments they are most comfortable with, without needing to learn a new proprietary language to access the features of the MQSS.
The modular design also allows for the extension to new and emerging frameworks.

\begin{figure*}[htbp]
    \vspace{10pt}
    \centering
    \includegraphics[width=0.96\textwidth]{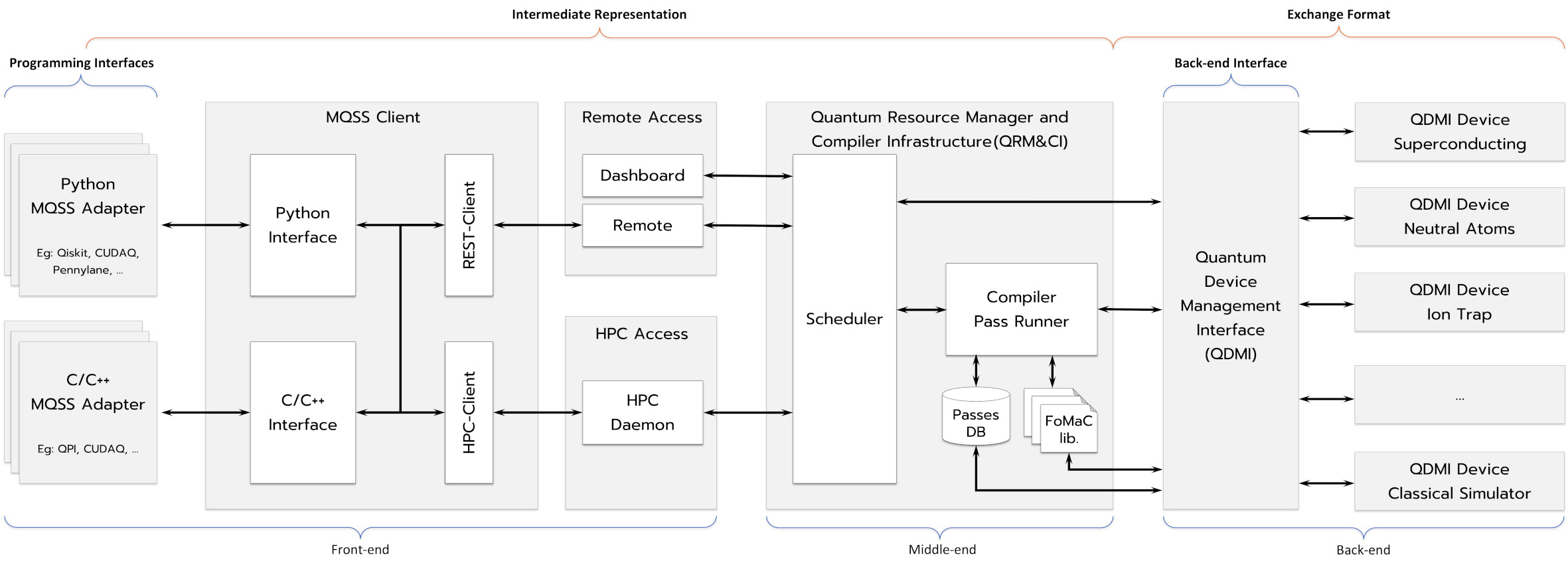}
    \caption{Low-level overview of the compilation and execution pipeline of the MQSS.}
    \Description{Detailed diagram showing the complete MQSS pipeline: job submission through front-end adapters, scheduling and device selection, device-agnostic compilation passes, device-specific compilation passes, code generation, execution via QDMI, and result handling back to the user.}
    \label{fig:overview-figure-low}
    \vspace{5pt}
\end{figure*}

We currently provide robust support for widely-used frameworks such as \emph{Qiskit}, \emph{PennyLane}, and \emph{CUDA-Q}.
This support allows the large communities built around these tools to seamlessly target MQSS-enabled back-ends, benefiting from the underlying compilation, optimization, and execution management capabilities.
Built on top of a uniform MQSS Client interface (center of \autoref{fig:overview-figure-low}), these adapters provide the final layer of abstraction, shielding the user from the complexities of the underlying hardware and software infrastructure.
The MQSS Client automatically detects whether a submission originates from within the HPC environment or from a remote location and routes it accordingly.
This also provides a hook for external tools, such as those at other data centers, to submit jobs to any given instance of the MQSS.

In addition to supporting the Python-based frameworks that are prominent in quantum computing, the MQSS offers programming models tailored to HPC environments, such as the Quantum Programming Interface (QPI;~\cite{kayaQPIProgrammingInterface2024}), enabling a more natural integration into native HPC languages, like C and C++, for developers accustomed to that ecosystem.

By supporting multiple adapters, the MQSS also enables users to take advantage of the distinctive strengths and specialized features offered by different libraries, thereby expanding the capabilities available to them.
For links and documentation to the available adapters, see:
\begin{itemize} %
    \item[\faBook] \small{\href{https://munich-quantum-software-stack.github.io/MQSS-Interfaces/} %
                               {https://munich-quantum-software-stack.github.io/ \newline MQSS-Interfaces}}
\end{itemize}

\section{Instantiations and Deployment}\label{sec:applications}

The MQSS has moved beyond conceptual design to a deployed, production-ready system.
Its architecture has been validated in a live HPC environment as part of the MQV initiative, where it serves as the unified access point to a heterogeneous set of quantum devices.
Operating within the production infrastructure of one of Europe's leading supercomputing centers for HPCQC, the stack has already been adopted by a diverse user base through multiple access modalities---from a user-friendly web portal to tightly-coupled hybrid access from within HPC jobs.
Experiences gathered during early user phases have provided valuable feedback for refining components and confirming that the stack can meet the demands of both research prototypes and real-world workloads.

To make this more tangible, a production instantiation is described next in the form of a representative end-to-end workflow, followed by a summary of a key study conducted during the early user phase.

\subsection{Production Instantiation at LRZ}
The MQSS is deployed at the Leibniz Supercomputing Centre (LRZ) as the single entry point to a heterogeneous set of quantum back-ends, including trapped-ion systems from AQT and superconducting systems from IQM.
In this setup, the MQSS components are deployed across different network zones to ensure security and manageability.
The user-facing components, such as the MQSS Dashboard (a web portal) and the MQSS Client endpoints, are accessible from the general internet and the HPC login nodes, respectively.
The middle-end layer of the stack, including the Quantum Resource Manager \& Compiler Infrastructure, runs in a dedicated, secured zone.
The QRM\&CI communicates with the various quantum systems via QDMI, which provides a standardized interface to the vendor-specific control software running in isolated, hardware-proximate environments.
This distributed deployment demonstrates the flexibility of the MQSS architecture, allowing it to be adapted to the specific security and networking requirements of a production HPC center.

\subsection{End-to-End Workflow through the MQSS}

Once installed and deployed, users can access the system in multiple ways that map directly onto the layers shown in \autoref{fig:overview-figure}.
The following representative end-to-end workflow demonstrates the interplay of these components in action, as shown in greater detail in \autoref{fig:overview-figure-low}.
The numbered steps below correspond to the labels in that figure.

\begin{figure*}[htbp]
    \vspace{10pt}
    \centering
    \includegraphics[width=0.85\textwidth]{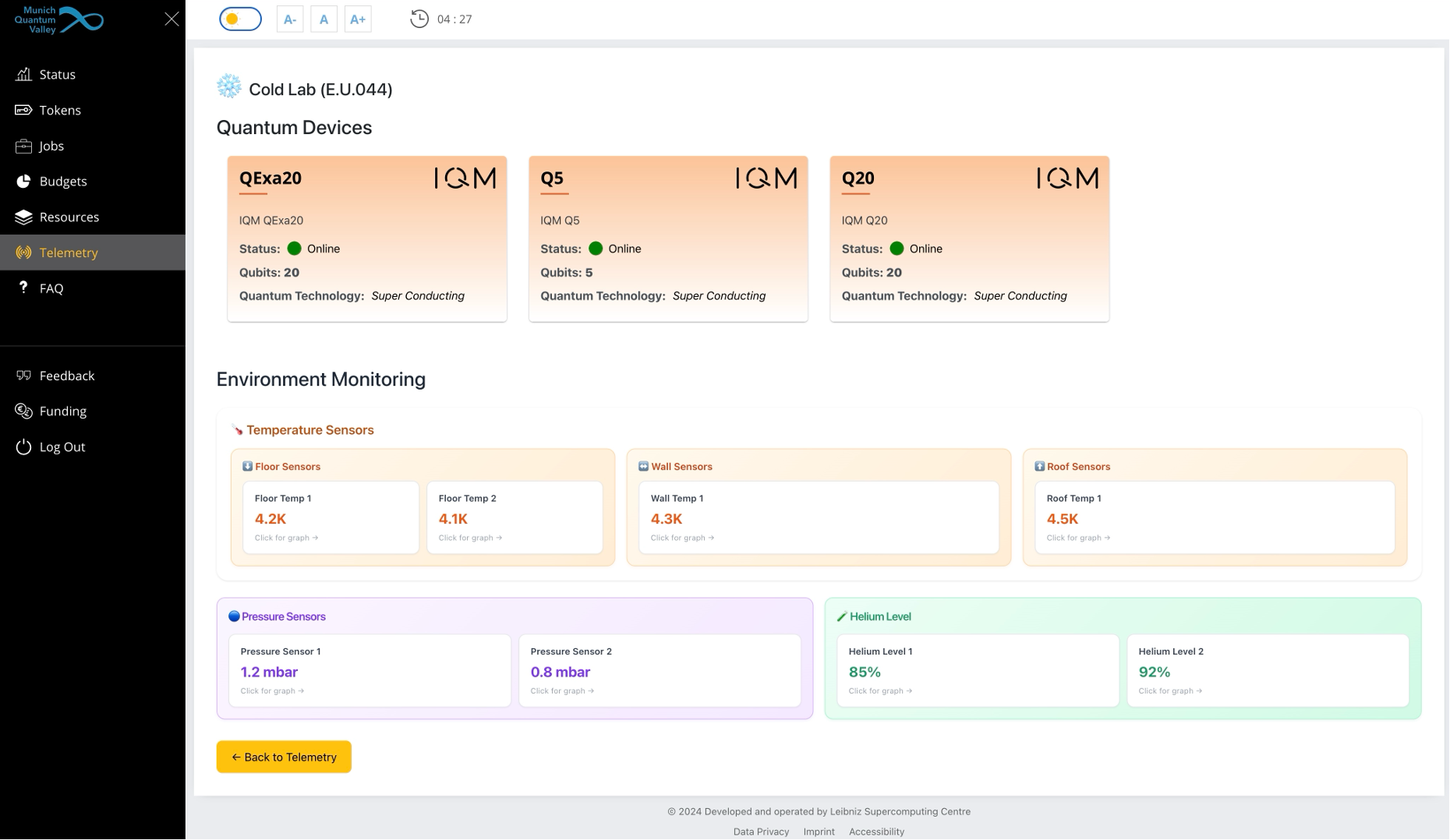}
  \caption{The MQSS Dashboard at LRZ, offering a web-based interface for job submission, monitoring, and management. Users can oversee token and budget usage, manage jobs and resources, and access detailed environment telemetry.}
  \Description{Screenshot of the MQSS Dashboard web interface at LRZ showing a user-friendly portal.}
  \label{fig:portal}
  \vspace{10pt}
\end{figure*}

Consider the scenario of a user who wishes to execute a hybrid quantum-classical algorithm.
\begin{enumerate}
    \item \textbf{Job Submission:} The user defines and implements their kernel in their favorite language or model and then submits it through corresponding Front-end Adapter. For interactive use, they might use a Python SDK, like Qiskit from a Jupyter notebook connected to the MQSS Dashboard (see \autoref{fig:portal}). For a large-scale hybrid workflow, they could code the kernel as a C++ application using QPI and execute that from a Slurm login node at LRZ. The MQSS Client receives the submission, handles authentication, and forwards the quantum part of the workflow to the QRM\&CI.
    \item \textbf{Scheduling and Device Selection:} The QRM\&CI's scheduler receives the quantum kernel. It first analyzes the circuit's requirements (e.g., qubit count, connectivity needs). Then, it queries the FoMaC Libraries for the current status and capabilities of all available QPUs. Based on this information and a user-defined (or default) policy, the scheduler performs a Pareto-optimal selection, identifying the most suitable back-end that balances various factors like queue time, expected fidelity, and execution speed. This integrated selection and scheduling step ensures that resources are allocated efficiently.
    \item \textbf{Device-Agnostic Compilation:} Once a set of candidate devices is known, the compiler infrastructure applies a series of device-agnostic optimization passes. A pass selector, potentially guided by machine learning heuristics, chooses transformations like gate cancellation, commutation-based reordering, or gate fusion to reduce the overall complexity of the circuit.
    \item \textbf{Device-Specific Compilation:} With a final target device selected by the scheduler, a second set of passes is applied. It uses detailed, live information from FoMaC Libraries (e.g., the native gate set, coupling map, or specific gate fidelities) to perform basis translation, qubit placement (mapping), and routing.
    \item \textbf{Code Generation:} The final, optimized circuit is lowered to an execution-ready representation. This could be a low-level IR, like QIR, or a vendor-specific format, potentially including pulse-level definitions, if supported by the back-end.
    \item \textbf{Execution:} The executable kernel is dispatched via QDMI to the selected back-end. QDMI manages the asynchronous execution and retrieves the results once the job is complete.
    \item \textbf{Result Handling:} The MQSS Client receives the raw measurement outcomes (e.g., bitstrings) from QDMI. It can perform initial post-processing, such as building histograms or applying measurement error mitigation, before returning the refined results and optional metadata (e.g., device used, calibration snapshot) to the user's application, again via the Front-end Adapter, enabling reproducibility and further analysis.
\end{enumerate}

The example illustrates the focus on extensibility and modularization, with each stage in the pipeline being policy-driven and replaceable.
Device-agnostic and device-specific pass selectors can be replaced or augmented with site-specific heuristics or ML models; mapping and routing passes can be tailored for different hardware technologies; scheduling policies can be adapted to specific operational priorities; and cost models can be tuned to optimize for metrics that matter in a given deployment.
Ultimately, this modularity ensures that, while the user experience and APIs remain stable, the underlying strategies for compilation and scheduling can evolve with hardware, workloads, and optimization objectives.

\subsection{Early-user Study}
As part of an early-user phase on the Q-Exa demonstrator, a hybrid quantum-classical workflow was executed through the MQSS on one of the 20-qubit superconducting devices available at LRZ~\cite{Mansfield_Integrating, bickley2025extendingquantumcomputingsubspace}.
The study targeted a chemically relevant problem using a multiscale Quantum Mechanics/Molecular Mechanics (QM/MM) approach, in which a quantum computation was embedded within a larger molecular simulation.
The workflow combined classical molecular dynamics and embedding methods on HPC nodes with quantum execution on the superconducting back-end, accessed seamlessly through the MQSS.

The problem was reduced to a 13-qubit Hamiltonian, enabling execution within the device's constraints.
On the QPU, a quantum algorithm was used to sample determinant subspaces, with the projected Hamiltonian solved classically.
Results achieved near chemical accuracy and consistently improved on baseline methods, demonstrating that embedded, noise-tolerant quantum algorithms can be effectively deployed within production HPC workflows.

This study illustrates how the MQSS supports tightly-coupled quantum acceleration in a real HPC environment, with device access, scheduling, and execution fully abstracted from the user's scientific problem while preserving flexibility in algorithmic and compilation strategies.

\section{Discussion / Related Work}\label{sec:comparison}
The development of quantum software stacks is a critical area of research and development worldwide, with several major initiatives pursuing different strategies.
The MQSS distinguishes itself through its emphasis on deep HPC integration, modularity, and open, community-driven standards, directly addressing the challenges of heterogeneity, integration, and the lack of practical community standards, as identified in \autoref{sec:why-stack}.

Major hardware vendors, such as \emph{IBM}, have developed comprehensive, vertically integrated software stacks (e.g., Qiskit~\cite{qiskit2024}) that are tightly coupled to their own hardware offerings.
While powerful and feature-rich, these stacks are naturally centered around a single vendor's ecosystem.
In contrast, the MQSS is designed from the ground up to be hardware- and vendor-agnostic, aiming to provide a unified interface to a diverse and heterogeneous landscape of quantum hardware from various providers.
Its modular architecture allows for the integration of any front-end framework or back-end system that adheres to its open interfaces.

National laboratories and other large-scale initiatives are also developing software solutions.
For example, the work at \emph{Oak Ridge National Laboratory} focuses on integrating quantum resources into their leadership computing facility~\cite{shehataBuildingSoftwareStack2025}.
While sharing the goal of HPCQC integration, the MQSS places a stronger emphasis on creating a portable, open-source stack that is not tied to a specific facility's infrastructure and can be deployed elsewhere.
Similarly, initiatives by organizations like the German Aerospace Center (\emph{DLR}) are building their own software platforms.
The MQSS differentiates itself through its long-standing, community-focused co-design process, involving a wide range of stakeholders from hardware physicists to domain-specific end-users, and its promotion of cross-platform standards, like QDMI, to foster a broader, more collaborative ecosystem.
Another related effort, the Quantum Resource Management Interface (QRMI;~\cite{sitdikov2025quantumresourcesresourcemanagement}), focuses primarily on standardized batch job submission to quantum systems, whereas QDMI provides a complete and comprehensive device management interface that also, next to its job management functionality, includes fine-grained, stateful session management and detailed, real-time query capabilities essential for dynamic compilation and optimization.

The MQSS's approach can be seen as a community-driven alternative to more monolithic or proprietary stacks.
By focusing on modularity, interoperability, and open standards, it aims at creating a flexible and sustainable software ecosystem that can adapt to the rapid evolution of quantum hardware and serve as a common platform for the entire HPCQC community, rather than a single organization or vendor.
This federated approach is crucial for tackling the challenges of integrating diverse hardware and software components, as highlighted in recent studies on the topic~\cite{humbleQuantumComputersHighPerformance2021}.

\newpage
\section{Conclusions and Outlook}\label{sec:conclusions}
In this paper, we have introduced the Munich Quantum Software Stack (MQSS), a comprehensive, open-source software ecosystem designed to bridge the gap between high-performance and quantum computing.
We have motivated the need for such a stack in enabling the vision of quantum acceleration, detailed the significant challenges in its development, and presented our community-driven approach to overcoming them.
The modular, multi-layered architecture of the MQSS, from the hardware-facing QDMI to the user-facing Front-end Adapters, provides a flexible and extensible platform for the development and execution of hybrid quantum-classical workflows.

The MQSS is not only a concept and academic prototype, but a tangible and evolving system, already deployed in a production HPC environment and gaining traction within the broader quantum community.
Our work demonstrates a viable path towards the deep integration of quantum processing units into classical supercomputing infrastructure, a critical step for unlocking the practical potential of quantum computing.

The future development of the MQSS will proceed along three main pillars.
First, \textbf{Stabilization}: we will continue to harden the existing components, improve interfaces, and provide robust support and documentation to maintain a production-ready, easy-to-use software stack.
Second, \textbf{Evolutionary expansion}: we will continuously expand the stack's capabilities, integrating new hardware back-ends, developing more sophisticated compiler optimizations, and enhancing support for NISQ-era applications. We will also intregrate community contributions towards an open ecosystem.
Third, \textbf{Revolutionary transformation}: we will proactively address the long-term challenges of fault-tolerant quantum computing, ensuring that the stack's architecture and interfaces are prepared for the paradigm shift to error-corrected quantum systems.
Through these efforts, the MQSS will remain at the forefront of quantum software, providing a vital platform for the future of integrated quantum and high-performance computing.

\begin{acks} %
This work was funded as part of the Munich Quantum Valley, which the Bavarian state government supports with funds from the Hightech Agenda Bayern via the Bavarian State Ministry for Science and Arts as well as the Bavarian State Ministry of Economics.
It further received funding from the German BMFTR via the grants of MUNIQC-SC, MUNIQC-Atoms, Euro-Q-Exa, Q-Exa and FullStaQD, the EU via the Quantum Flagship projects OSQ+ and Millenion, the European Research Council (ERC) under the European Union's Horizon 2020 research and innovation program (grant agreement No. 101001318), and EuroHPC via Euro-Q-Exa.
Additionally funding was provided by the State of Upper Austria in the frame of the COMET program (managed by the FFG) and by NSF grant CCF-2313083, as well as via industrial support by BMW.
\end{acks}

\newpage
\clubpenalty=10000
\widowpenalty=10000
\interlinepenalty=10000
\bibliographystyle{ACM-Reference-Format}
\bibliography{header,lit}

\end{document}